\documentclass[12pt]{article}
\usepackage{graphicx}
\usepackage{cite}
\usepackage{color}
\begin{document}
\title{The $X(4260)$ and possible confirmation\\
of
$\psi(3D)$, $\psi(5S)$, $\psi(4D)$, $\psi(6S)$ and $\psi(5D)$
in $J/\psi\pi\pi$}
\author{
Eef~van~Beveren$^{\; 1}$ and George~Rupp$^{\; 2}$\\ [10pt]
$^{1}${\small\it Centro de F\'{\i}sica Computacional,
Departamento de F\'{\i}sica,}\\
{\small\it Universidade de Coimbra, P-3004-516 Coimbra, Portugal}\\
{\small\it eef@teor.fis.uc.pt}\\ [10pt]
$^{2}${\small\it Centro de F\'{\i}sica das Interac\c{c}\~{o}es Fundamentais,}\\
{\small\it Instituto Superior T\'{e}cnico, Universidade T\'{e}cnica de
Lisboa,}\\
{\small\it Edif\'{\i}cio Ci\^{e}ncia, P-1049-001 Lisboa, Portugal}\\
{\small\it george@ist.utl.pt}\\ [.3cm]
{\small PACS number(s): 14.40.Gx, 13.66.Bc, 14.40.Lb, 14.20.Lq, }
}


\maketitle
\begin{abstract}
Data on $e^+e^-\to J/\psi\pi\pi$ and the $X(4260)$ enhancement by
the BABAR collaboration \cite{PRL95p142001,ARXIV08081543} are analysed by
modelling missing signals that should be present in the complete production
amplitude with vector quantum numbers in the charmonium region. Thus, it is
shown that the data contain evidence for the existence of the $\psi(5S)$, the
$\psi(4D)$, the $\psi(6S)$ and the $\psi(5D)$ $c\bar{c}$ vector states, and
furthermore a clear indication for the mass and width of the $\psi(3D)$.
Moreover, it is shown that signs of the $\psi(3S)$, $\psi(2D)$
and $\psi(4S)$ can be observed in the same data.
Finally, it is argued that the $X(4260)$ enhancement is not a resonance, but
rather a phenomenon connected with the opening of the
$D_{s}^{\ast}D_{s}^{\ast}$ threshold and the coupling to the
$J/\psi f_0(980)$ channel.
\end{abstract}

In Ref.~\cite{PRL95p142001}, the BABAR collaboration announced the
observation of a new vector state in the charmonium region, originallly
baptised as $Y(4260)$ but now included in the PDG \cite{PLB667p1} tables as
$X(4260)$, by studying the $e^{+}e^{-}\to J/\psi\pi^{+}\pi^{-}$ cross section.
The experimental analysis resulted in a mass and width
for this enhancement of $M=(4259\pm 8)$(stat.)$^{+2}_{-6}$(sys.)~MeV
and $\Gamma_{\mathrm{tot}}=88\pm 23$(stat.)$^{+6}_{-4}$(sys.)~MeV,
\cite{PRL95p142001}, respectively, with a significance of 5--7 $\sigma$.
The BABAR result, later confirmed and also seen
in $\pi^{0}\pi^{0}J/\psi$ as well as $K^{+}K^{-}J/\psi$
by the CLEO collaboration \cite{PRL96p162003},
whereas the BELLE collaboration \cite{PRL99p182004} 
confirmed a similar structure
in $J/\psi\pi^{+}\pi^{-}$
has been studied in a variety of theoretical models \cite{PLB625p212},
namely as a standard vector charmonium state ($4S$)
\cite{PRD72p031503},
a mesonic or baryonic molecule
\cite{PRD72p054023},
a gluonic excitation (hybrid)
\cite{PLB631p164},
or a $cq\bar{c}\bar{q}$ tetraquark
\cite{PRD72p031502}.

A peculiar aspect of this experimental observation
is that the main signal of $e^{+}e^{-}\to\pi^{+}\pi^{-}J/\psi$
does not seem to show any sign of the known vector charmonium states.
In Ref.~\cite{PRL95p142001} of the BABAR collaboration,
one reads:
``{\it no other structures are evident at the masses
of the quantum number} $J^{PC}=1^{--}$ {\it charmonium states,
i.e., the} $\psi(4040)$, $\psi(4160)$, {\it and} $\psi(4415)$''.
This at first sight very surprising observation has not attracted
due attention in theoretical approaches to the $X(4260)$, which focus
exclusively on trying to model this structure, but forget about the
established yet not visible $c\bar{c}$ resonances. Therefore, our strategy
here will be different, attempting to understand both aspects of the BABAR
data, namely what {\em is seen} \/and what {\em is not seen}.

In  Refs.~\cite{HEPPH0605317} and \cite{ARXIV08111755}, we discussed
signs of the $\psi(4160)$ and $\psi(4415)$, respectively,
in BABAR data. Here, we shall follow a similar strategy,
since with the presence of many charmed-pair thresholds it is an
extremely difficult and, moreover, not very transparent
task to precisely fit the data. We shall assume that the reaction of
electron-positron annihilation into multi-hadron final states basically
takes place via one photon, hence with $J^{PC}=1^{--}$ quantum numbers.
Consequently, when the photon materialises into a pair of current quarks,
which couple via the quark-antiquark propagator to the final multi-hadron
state, we may assume that the intermediate propagator carries the quantum
numbers of the photon. Moreover, alternative processes are suppressed.

Furthermore, since $J/\psi$ contains
a dominant contribution of charm-anticharm, we assume that in the reaction
$e^{+}e^{-}\to J/\psi\pi^{+}\pi^{-}$ initially a $c\bar{c}$ pair is formed.
The subsequent formation of the pion pair, being OZI \cite{OZI} forbidden,
must be a relatively slow process, hence easily superceded
by decay of the $c\bar{c}$ pair into pairs of charmed hadrons,
via the formation of a light quark-antiquark pair.
The latter process will in particular deplete
the $c\bar{c}$ propagator
at the opening of thresholds of charmed hadron pairs
and at $c\bar{c}$ resonances,
resulting in dips in the $J/\psi\pi^{+}\pi^{-}$ signal.
In the following, we shall search for such dips
and show that the missing signal can very well be
identified with the known, and also some new \cite{EPL85p61002},
vector charmonium resonances.

\begin{figure}[htbp]
\begin{center}
\begin{tabular}{cc}
\scalebox{0.78}{\includegraphics{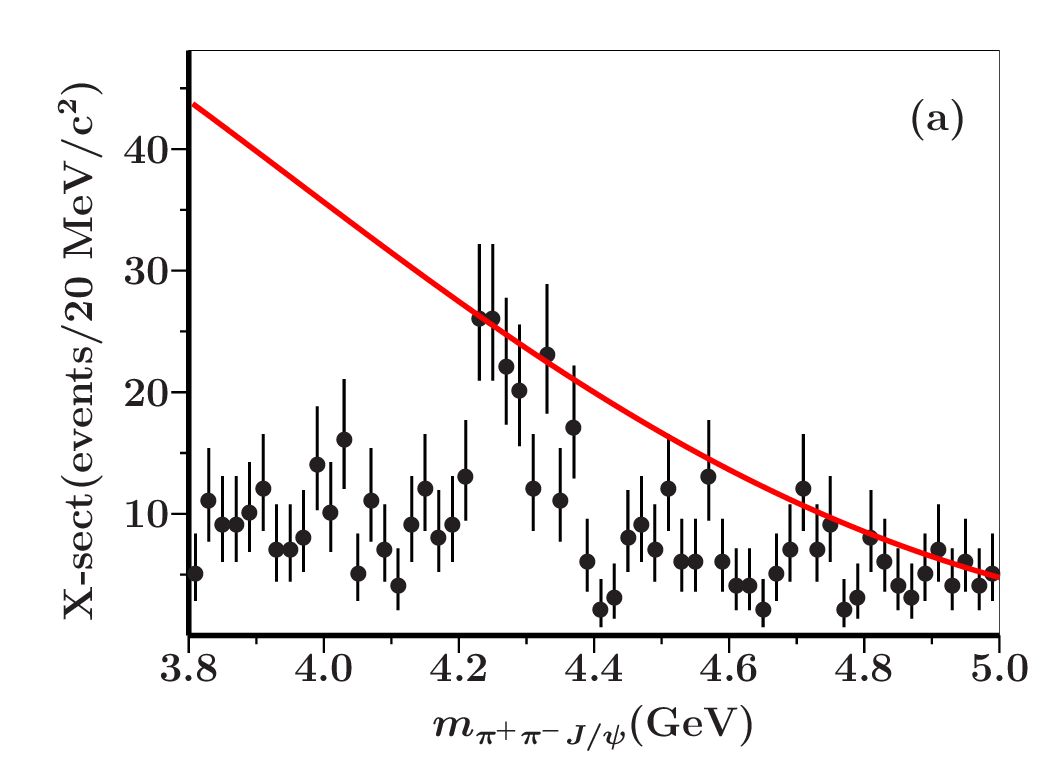}} &
\scalebox{0.78}{\includegraphics{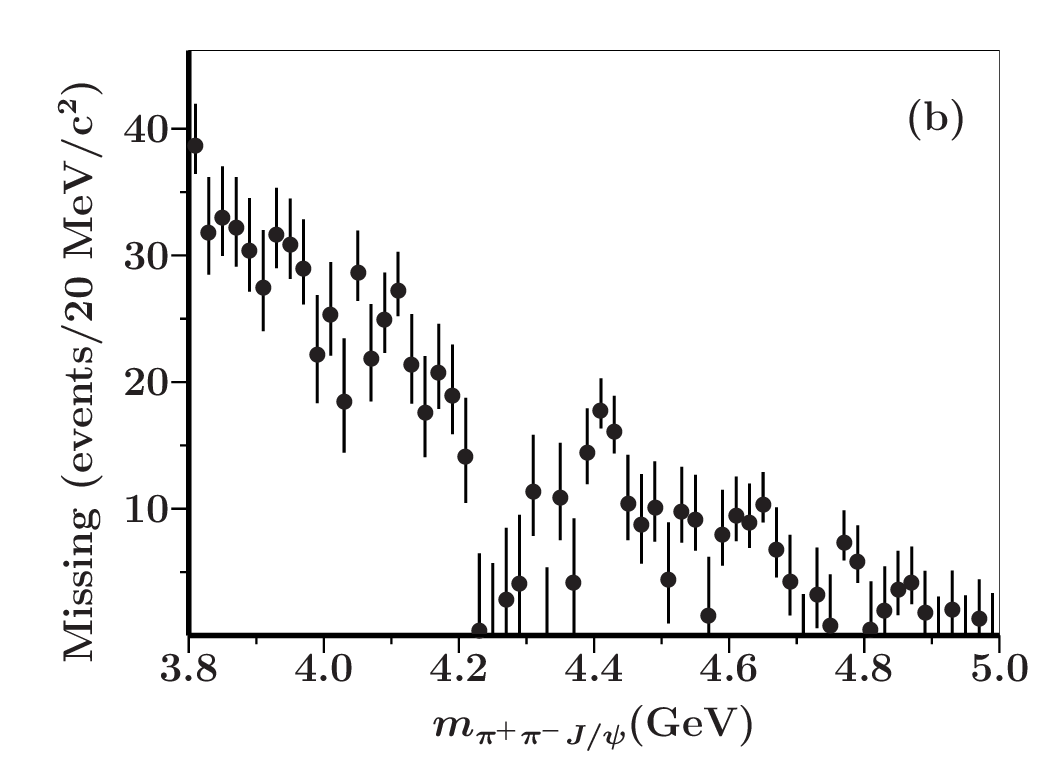}}\\ [-15pt]
\end{tabular}
\end{center}
\caption[]{\small
(a): Experimental data
for $e^{+}e^{-}$ annihilation into $J/\psi\pi^{+}\pi^{-}$
from the BABAR collaboration \cite{PRL95p142001},
and the theoretical line shape of production \cite{ARXIV08111755}.\\
(b): The signal that is missing in the BABAR data
with respect to the line shape in (a).
}
\label{bled}
\end{figure}

In Ref.~\cite{ARXIV08111755}, we proposed
a genuine $J/\psi\pi^{+}\pi^{-}$ production amplitude
as shown in Fig.~\ref{bled}a.
The missing-signal amplitude, shown in Fig.~\ref{bled}b,
is determined by the difference
of the true theoretical amplitude and the actual data.
However, analysing the result of Fig.~\ref{bled}b
is not an easy task at energies below 4.2 GeV,
since many different processes have accumulating effects.
This is unnecessary though.  We may very well restrict us to the
visible signal.
The reason is that for the theoretical curve of Fig.~\ref{bled}a
we assumed in Ref.~\cite{ARXIV08111755} that no further processes
occur but just the reaction $e^{+}e^{-}\to J/\psi\pi^{+}\pi^{-}$.
Further on in the latter paper, we discussed
how the various interactions deform this no-interaction
prediction towards the observed data.
Here, we assume the existence of such interactions from the start.

\begin{figure}[htbp]
\begin{center}
\begin{tabular}{cc}
\scalebox{0.78}{\includegraphics{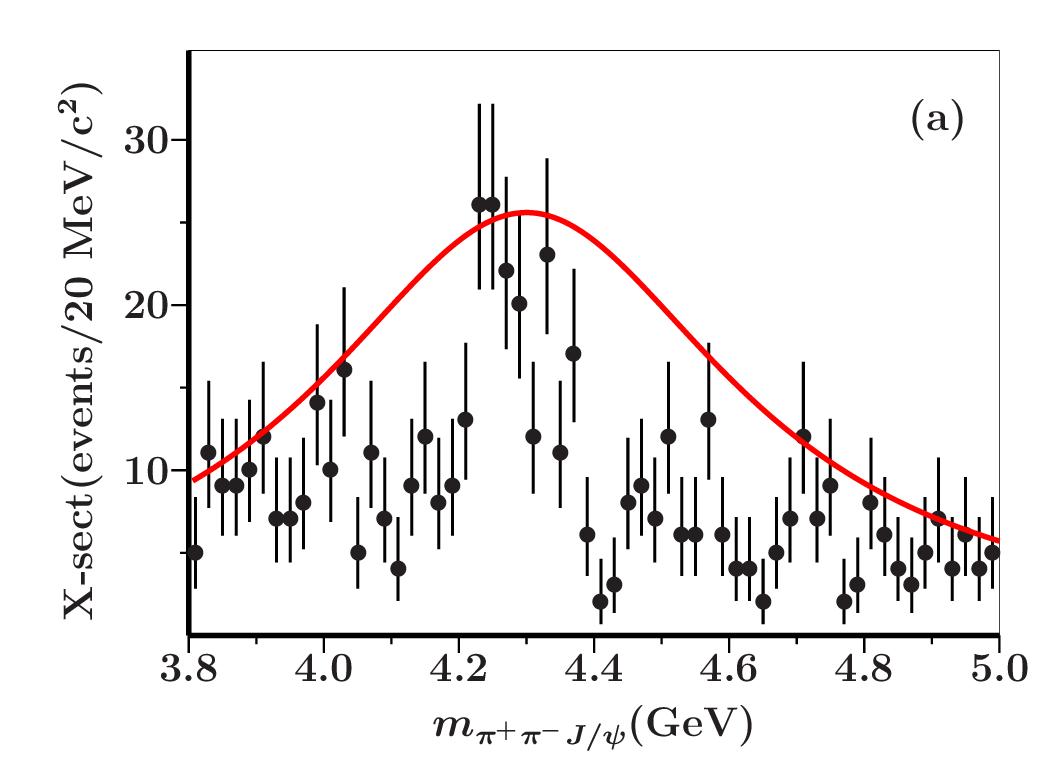}} &
\scalebox{0.78}{\includegraphics{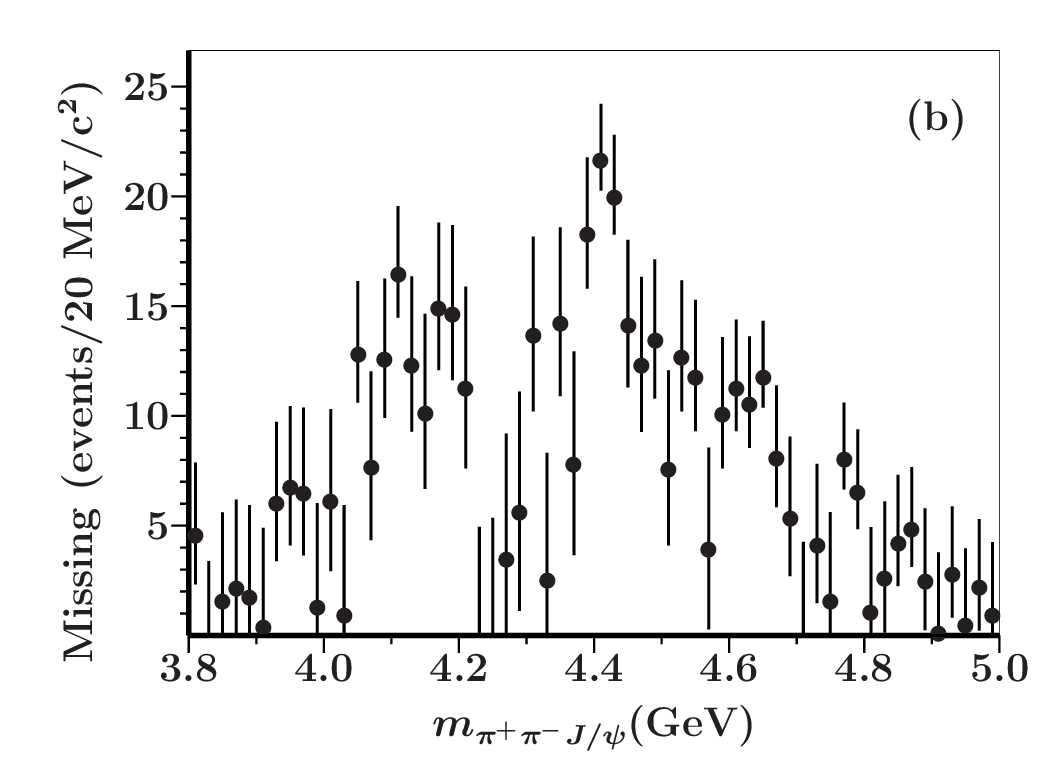}}\\ [-15pt]
\end{tabular}
\end{center}
\caption[]{\small
(a): Experimental data
for $e^{+}e^{-}$ annihilation into $J/\psi\pi^{+}\pi^{-}$
from the BABAR collaboration \cite{PRL95p142001},
and the line shape of a broad structure
with a width of 750 MeV and a central value of 4.3 GeV.\\
(b): The signal missinging in the BABAR data
with respect to the line shape in (a).
}
\label{babar}
\end{figure}

\begin{figure}[htbp]
\begin{center}
\begin{tabular}{cc}
\scalebox{0.78}{\includegraphics{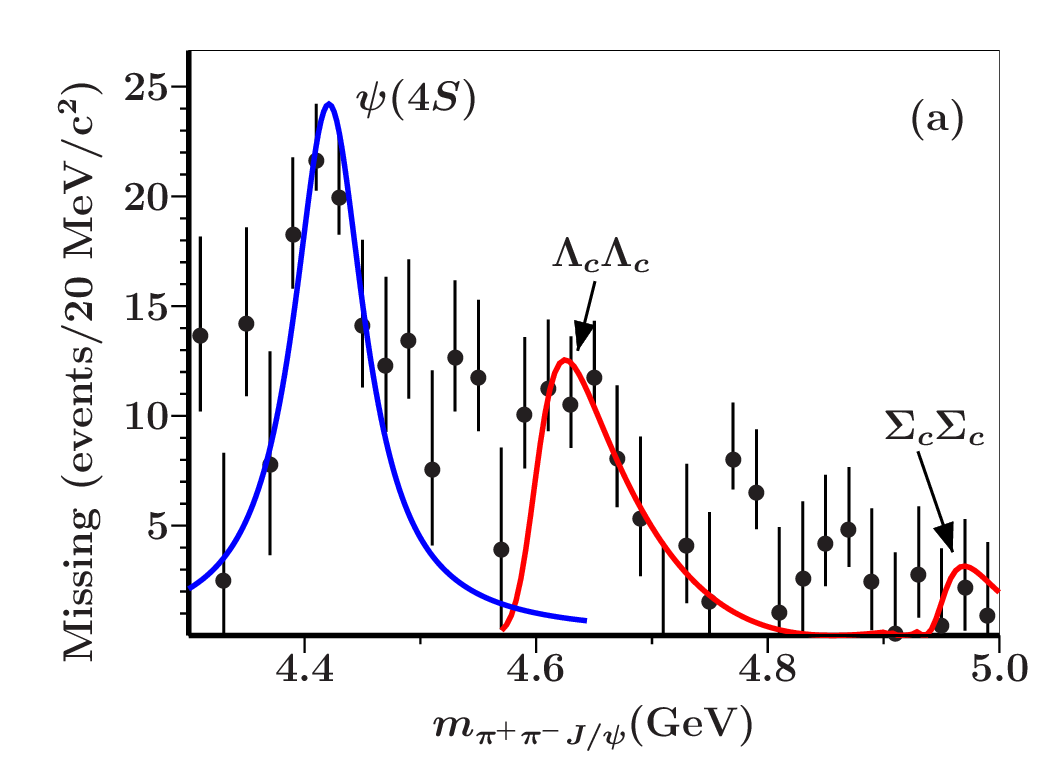}} &
\scalebox{0.78}{\includegraphics{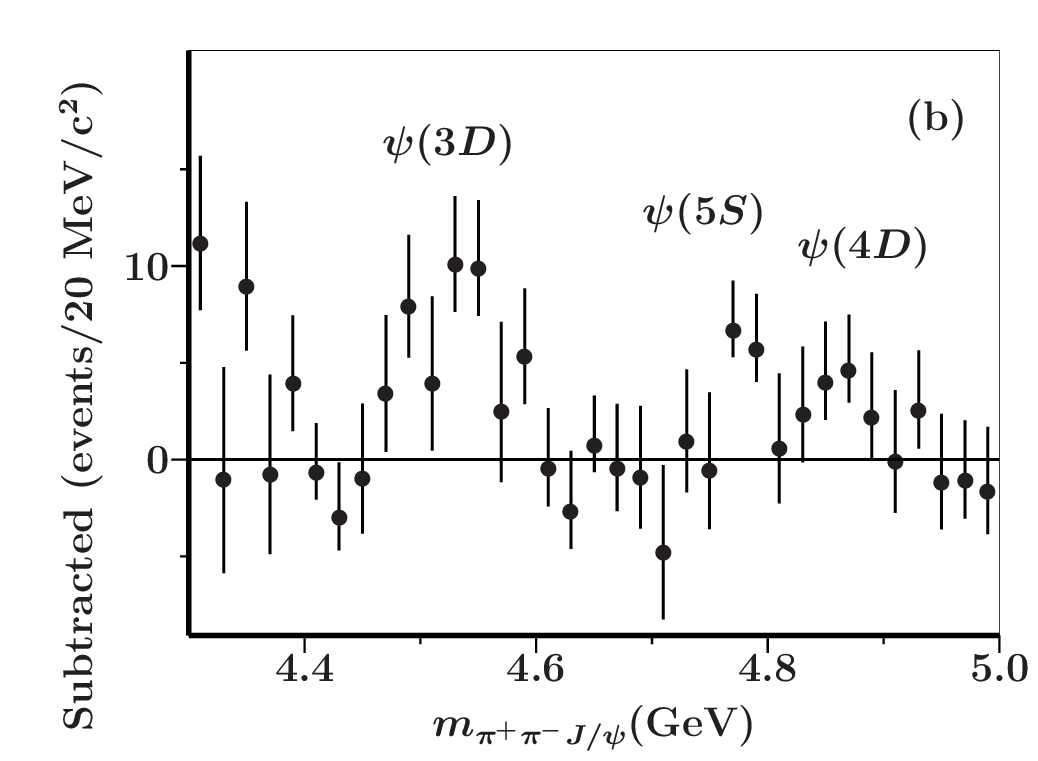}}\\ [-15pt]
\end{tabular}
\end{center}
\caption[]{\small
(a): The missing signal
in $e^{+}e^{-}$ annihilation into $J/\psi\pi^{+}\pi^{-}$
compared to a Breit-Wigner approximation for
the $\psi (4S)$ resonance at 4.421 GeV
with a width of 75 MeV (blue curve),
and to the line shapes for the processes
$e^{+}e^{-}\to\Lambda_{c}\Lambda_{c}$
and $e^{+}e^{-}\to\Sigma_{c}\Sigma_{c}$,
which have thresholds at 4.572 GeV and 4.907 GeV, respectively.\\
(b): The remaining signal after removing the line shapes
of (a) from the data.
}
\label{missing}
\end{figure}

Thereto, we construct an envelope for the data
via a Breit-Wigner shape covering them.
This is shown in Fig.~\ref{babar}a.
The line shape of the broad envelope
has a maximum at 4.3 GeV and a width of 750 MeV.
In Fig.~\ref{babar}b we then display the missing-signal data,
which are obtained by subtracting the BABAR data
from the theoretical line shape.
The latter signal must be due to $e^{+}e^{-}$ annihilation processes
into channels other than $J/\psi\pi^{+}\pi^{-}$.
For example, the large structure in the center of Fig.~\ref{babar}b,
just above 4.4 GeV is readily recognised as the $\psi (4415)$,
which dominantly decays into charmed pairs of mesons,
hence eating away signal from $J/\psi\pi^{+}\pi^{-}$.
In Fig.~\ref{missing}a, we show the Breit-Wigner approximation
for a resonance at 4.421 GeV \cite{PLB667p1}
and a width of 75 MeV (our estimate),
which indeed fits the missing signal quite well.

However, there is more in this energy region.
At about 4.57 GeV, $e^{+}e^{-}$ can annihilate into a pair of charmed
baryons, namely $\Lambda_{c}\Lambda_{c}$,
thus eating away signal from $J/\psi\pi^{+}\pi^{-}$.
In Ref.~\cite{EPL85p61002}, we and our co-authors
determined the line shapes of the reactions
$e^{+}e^{-}\to\Lambda_{c}\Lambda_{c}$
and $e^{+}e^{-}\to\Sigma_{c}\Sigma_{c}$.
This is also shown in Fig.~\ref{missing}a.
It fits well the present missing-signal data.

When we  next determine the signal that is left after
subtracting the line shapes of the $\psi (4S)$ resonance, as well as
the $e^{+}e^{-}\to\Lambda_{c}\Lambda_{c}$ and
$e^{+}e^{-}\to\Sigma_{c}\Sigma_{c}$ processes,
we find the result depicted in Fig.~\ref{missing}b.
It shows a clear enhancement around 4.53 GeV,
which is probably the $\psi (3D)$ resonance,
implicitly predicted by us, in collaboration with C.~Dullemond,
in Ref.~\cite{PRD21p772},
and explicitly by S.~Godfrey and N.~Isgur in Ref.~\cite{PRD32p189}.
At higher energies, we observe two feable enhancements, which we interprete
as the $\psi (5S)$, close to 4.79 GeV, and the $\psi (4D)$, at about
4.87 GeV. In a our recent analysis in Ref.~\cite{EPL85p61002}, we obtained
the latter two resonances at precisely these masses, directly in the
$\Lambda_{c}\Lambda_{c}$ data  published by the BELLE collaboration
\cite{PRL101p172001}, and we deduced the existence of the former resonance
from the behaviour at threshold. The fact that our procedure here works is
largely due to the very precise BABAR data \cite{PRL95p142001}.
Moreover, it is quite reassuring that our alternative strategies of analysis
in Ref.~\cite{EPL85p61002} and the present paper, on the basis of data from
two different experimental collaborations, result in coinciding predictions.

\begin{figure}[htbp]
\begin{center}
\begin{tabular}{cc}
\scalebox{0.78}{\includegraphics{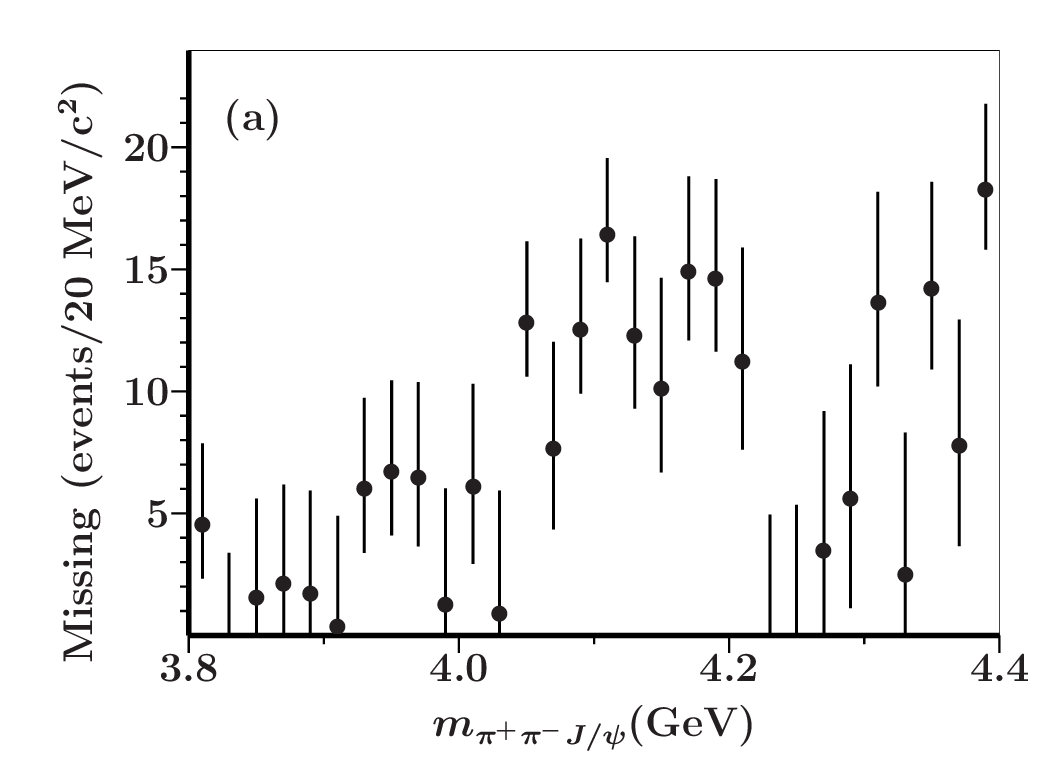}} &
\scalebox{0.78}{\includegraphics{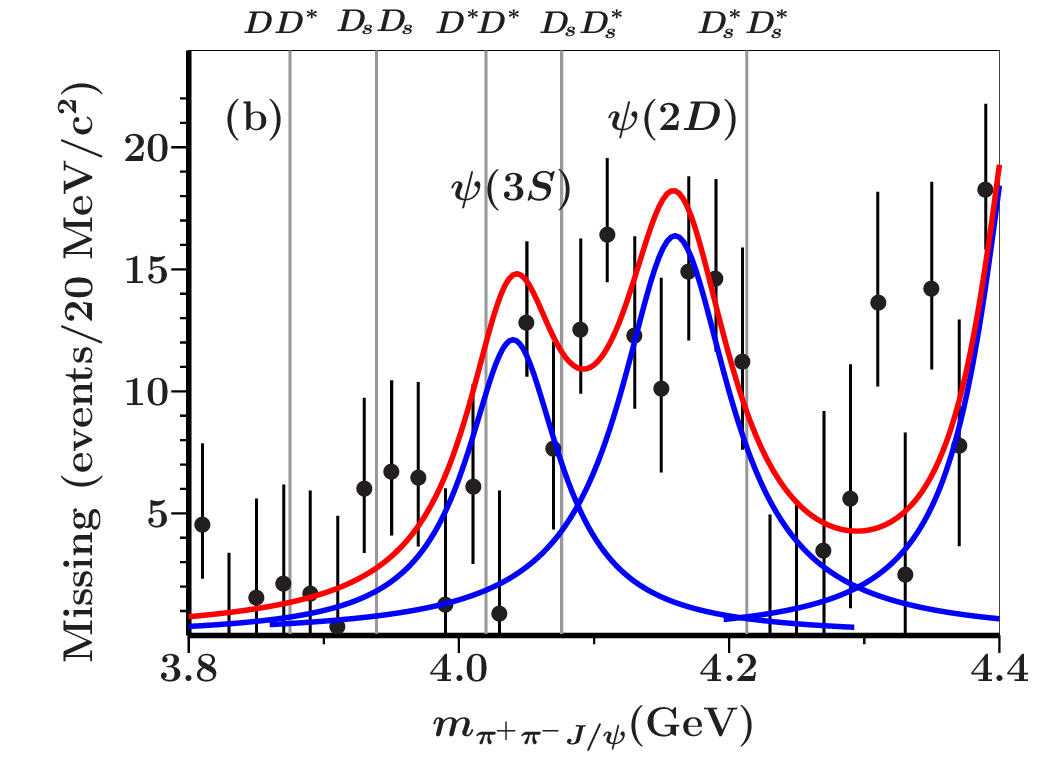}}\\ [-15pt]
\end{tabular}
\end{center}
\caption[]{\small
(a): Missing-signal data
for $e^{+}e^{-}$ annihilation into $J/\psi\pi^{+}\pi^{-}$
in the invariant-mass region 3.8--4.4 GeV, extracted from Fig.~\ref{babar}b;\\
(b): The $\psi(4040)$, $\psi(4160)$, and tail of the $\psi(4415)$
Breit-Wigner structures.
The vertical lines represent the masses
of the various charmed-meson-pair thresholds.
The upper curve, in red, shows the sum of the squares of the
amplitudes.
}
\label{babarlow}
\end{figure}

At the lower side of the energy spectrum, we expect to see two charmonium
vector states, namely the $\psi(4040)$ and the $\psi(4160)$,
and furthermore the opening of several open-charm channels, viz.\
$DD^{\ast}$ at 3.875 GeV, $D^{\ast}D^{\ast}$ at 4.02 GeV, $D_{s}D_{s}$ at
3.939 GeV, $D_{s}D_{s}^{\ast}$ at 4.076 GeV, and $D_{s}^{\ast}D_{s}^{\ast}$
at 4.213 GeV. In Fig.~\ref{babarlow}a we display the missing-signal data
for energies ranging from 3.8 GeV to 4.4 GeV, whereas in Fig.~\ref{babarlow}b
we show what may be expected in that invariant-mass region
for the $\psi(4040)$ and $\psi(4160)$ resonances.

We observe that the missing data are well filled up with precisely these two
resonances.  Nevertheless, just summing up Breit-Wigner structures
is, of course, not the correct strategy for a detailed analysis of the data.
Moreover, we have not included the opening of the open-charm production
thresholds. The least one may expect is some interference effects between the
$\psi(4040)$ and $\psi(4160)$ resonances. This has been studied in
Ref.~\cite{HEPPH0605317} and will be repeated in the following.

The data of Ref.~\cite{PRL95p142001}, shown in Figs~\ref{bled}a and
\ref{babar}a, indeed seem to indicate interference
between the $\psi(4040)$ and $\psi(4160)$, since a sequence of 8 data points
behave exactly as expected for a resonance, i.e., the $\psi(2D,4160)$
in the tail of another, lower-mass resonance, viz.\ the $\psi(3S,4040)$.
We are well aware that the authors of Ref.~\cite{PRL95p142001}
did \em not \em \/see this feature in their data. Nevertheless, we are
convinced the $\psi(2D,4160)$ structure is there.
In order to make our
\begin{figure}[htbp]
\begin{center}
\begin{tabular}{cc}
\includegraphics[height=150pt, angle=0]{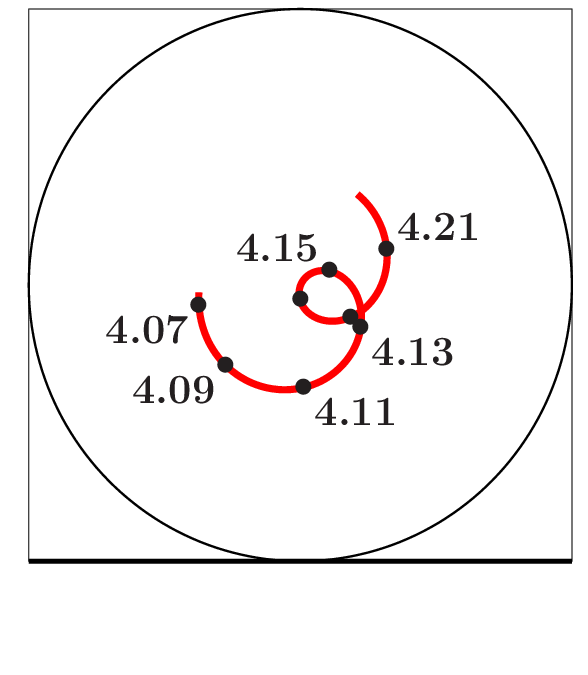} &
\includegraphics[height=150pt, angle=0]{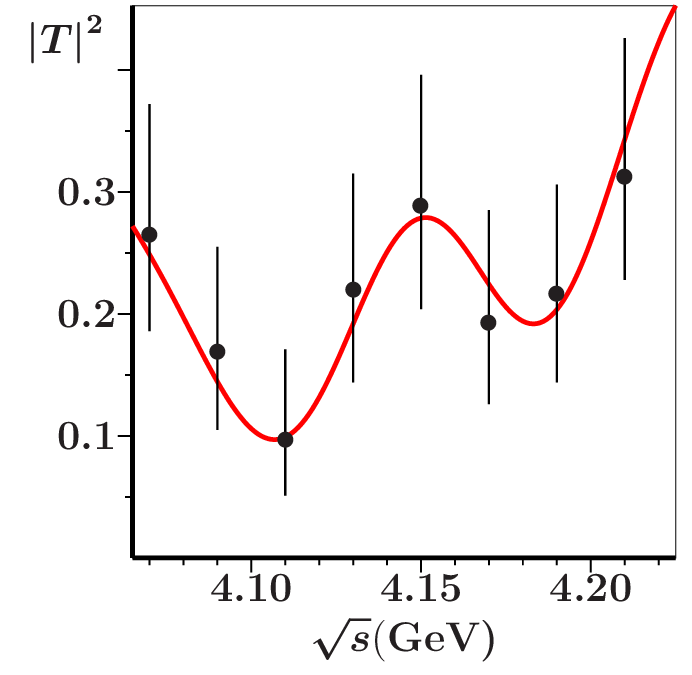} \\[-8mm]
\end{tabular}
\end{center}
\caption[]{\small Simulated phase motion around the $\psi(2D,4160)$ (left);
corresponding cross section, with a sequence of 8 data points of
Ref.~\cite{PRL95p142001} (right).}
\label{psi2D}
\end{figure}
point, assuming reasonable values for the amplitude, we simulate in
Fig.~\ref{psi2D} a possible phase motion that is compatible with the
mentioned, and also shown, 8 data points. We repeat, the depicted phase
motion is just a simulation, and \em not \em \/a prediction of our model.

To complete our analysis, we also have a look at new though preliminary
BABAR data \cite{ARXIV08081543} for $e^{+}e^{-}\to J/\psi\pi^{+}\pi^{-}$.
These exhibit a much more pronounced peak in the $X(4260)$ region,
and, moreover, a rather constant signal for the remaining invariant masses.
Here, we shall inspect the structure of the background for the preliminary 
BABAR data in the invariant-mass region 4.8--5.4 GeV.
\begin{figure}[htbp]
\begin{center}
\begin{tabular}{cc}
\scalebox{0.78}{\includegraphics{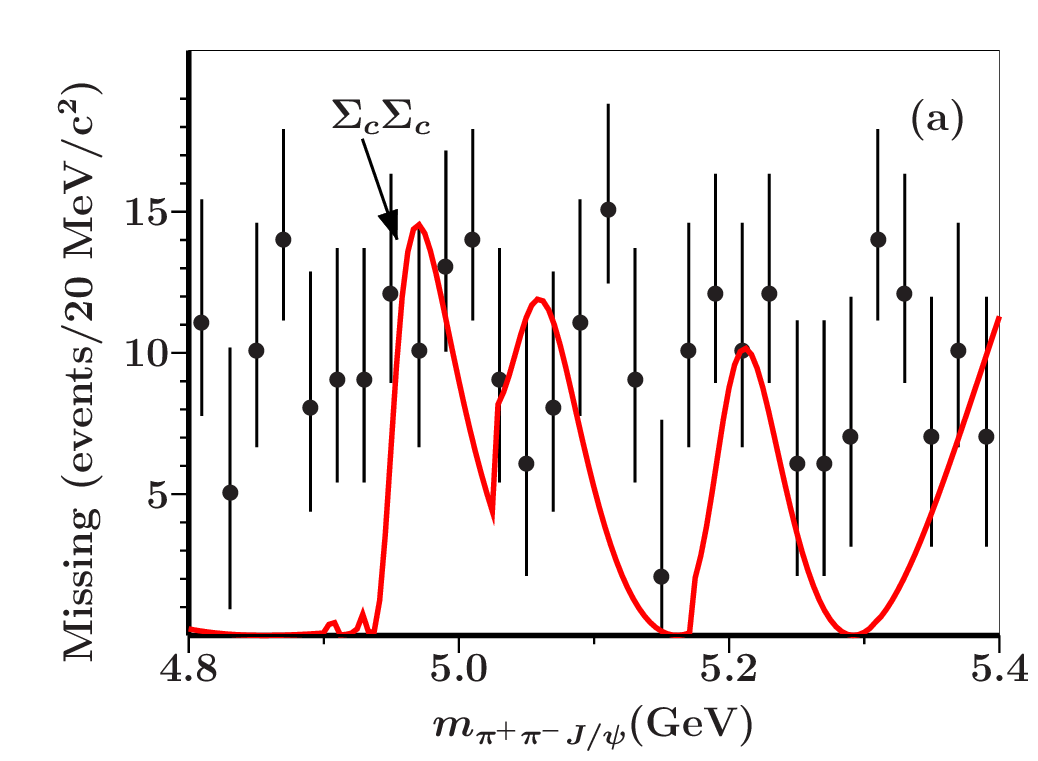}} &
\scalebox{0.78}{\includegraphics{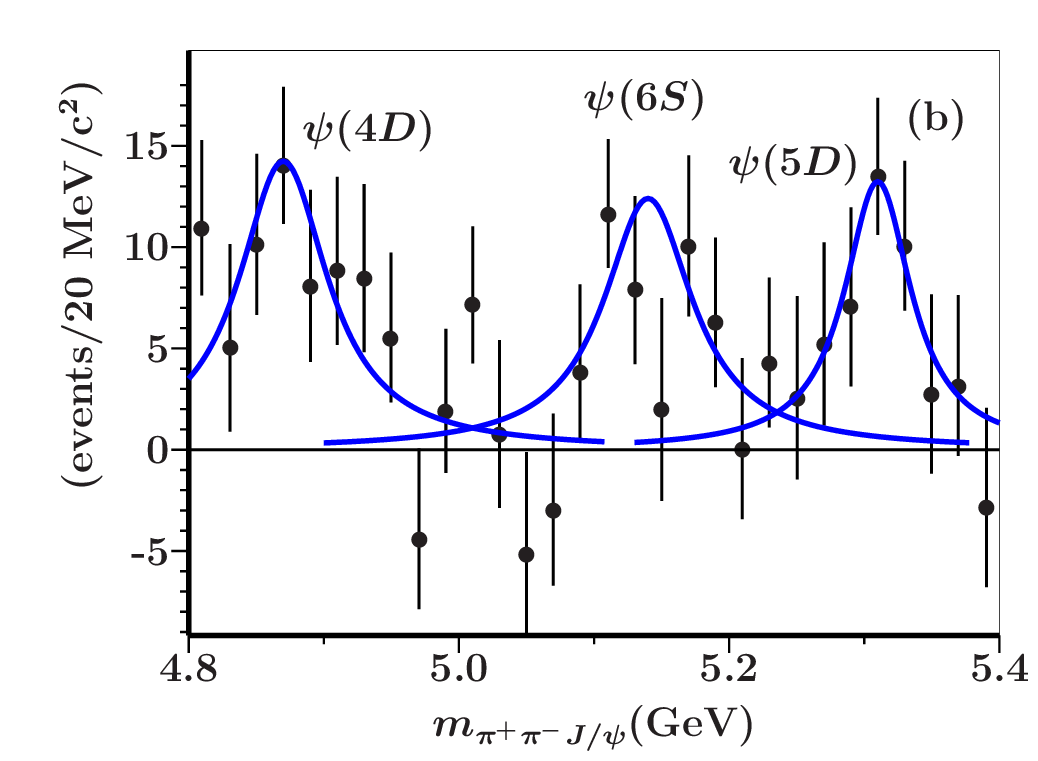}}\\ [-15pt]
\end{tabular}
\end{center}
\caption[]{\small
(a): Missing-signal data
for $e^{+}e^{-}$ annihilation into $J/\psi\pi^{+}\pi^{-}$
in the invariant-mass region 4.8--5.4 GeV,
extracted from preliminary BABAR data \cite{ARXIV08081543}.
The solid line (red), representing the signal for
$e^{+}e^{-}\to\Lambda_{c}\Lambda_{c}$,
and which shows here the opening of the various
$\Sigma_{c}\Sigma_{c}$ channels,
is taken from Ref.~\cite{EPL85p61002}.\\
(b): The remaining signal, after subtraction of the solid line
in (a), with the $\psi(4D)$, $\psi(6S)$ and the $\psi(5D)$ as
Breit-Wigner structures.
}
\label{babarprl}
\end{figure}
In Fig.~\ref{babarprl}a we show the BABAR data, but upside down and shifted
22 events upwards, as a constant envelope comprising 22
events seems good enough in the relevant invariant-mass region.
Furthermore, we display in Fig.~\ref{babarprl}a the signal for the reaction
$e^{+}e^{-}\to\Lambda_{c}\Lambda_{c}$, taken from Ref.~\cite{EPL85p61002},
and which shows here the opening of the
$\Sigma_{c}(2455)\bar{\Sigma}_{c}(2455)$ channel and also of higher thresholds
involving charmed $\Sigma$ baryons.
In Fig.~\ref{babarprl}b, we depict the remaining signal,
after subtracting the solid line of Fig.~\ref{babarprl}a.
We clearly observe the $\psi(4D)$ $c\bar{c}$ resonance, and find additional
indications for the $\psi(6S)$ and $\psi(5D)$ resonances.

\begin{figure}[htbp]
\begin{center}
\begin{tabular}{cc}
\scalebox{0.78}{\includegraphics{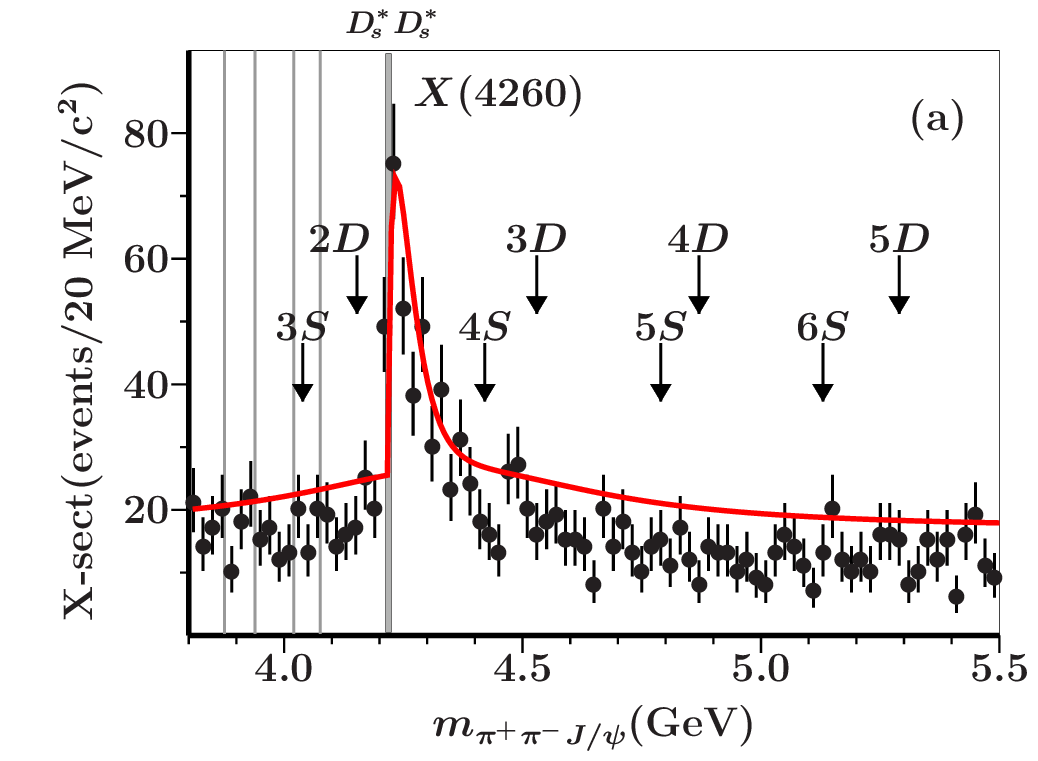}} &
\scalebox{0.78}{\includegraphics{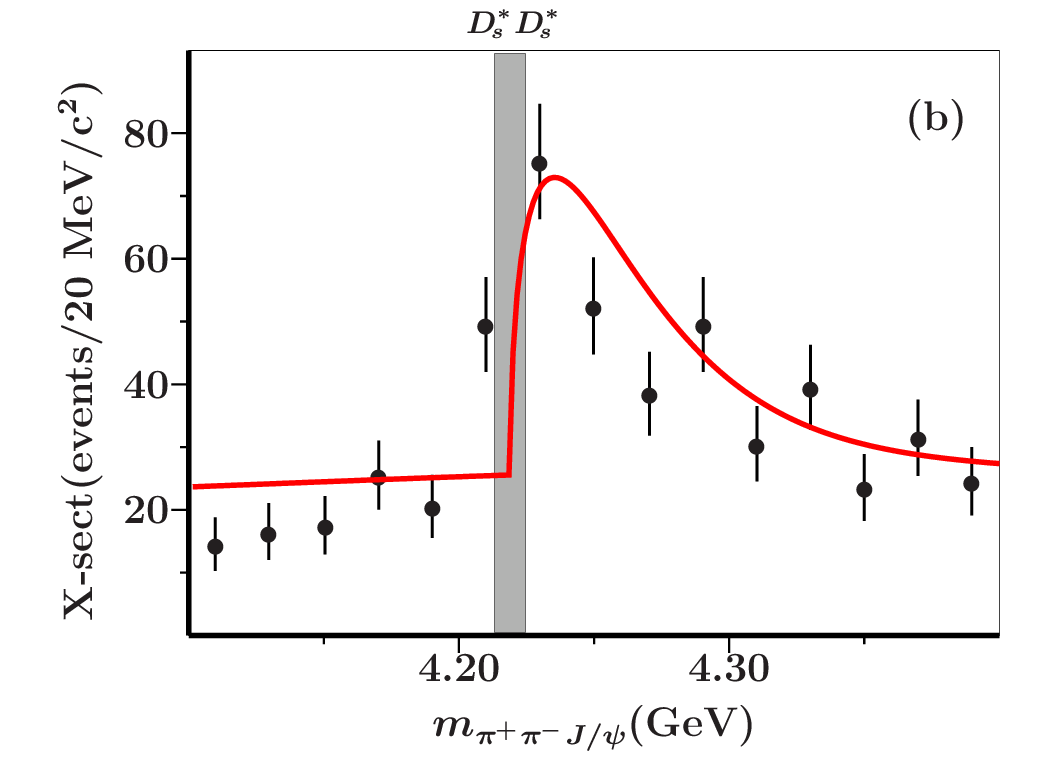}}\\ [-15pt]
\end{tabular}
\end{center}
\caption[]{\small
(a): The new preliminary BaBar data \cite{ARXIV08081543}
for $e^{+}e^{-}$ annihilation into $J/\psi\pi^{+}\pi^{-}$
in the invariant-mass regions 3.8--5.5 GeV (a) and 4.1--4.4 GeV (b).
The solid line comprises three contributions:
a constant background of 17 events, a bell-shaped Breit-Wigner-like
contribution, centred at 4.35 GeV and with a width of 750 MeV,
and an eye-guiding line through the data around the $X(4260)$ peak.
In (a) we indicate the central masses of the various
$c\bar{c}$ resonances discussed in this work,
and furthermore the threshold positions of
$DD^{\ast}$, $D^{\ast}D^{\ast}$,
$D_{s}D_{s}$, $D_{s}D_{s}^{\ast}$, and $D_{s}^{\ast}D_{s}^{\ast}$
(vertical lines).
In (b) we depict a possible spreading of the $D_{s}^{\ast}D_{s}^{\ast}$
threshold position in accordance with different experiments (see text).
}
\label{babarnew}
\end{figure}

In Fig.~\ref{babarnew}a we show the new preliminary BABAR data just as they
are and, in Fig.~\ref{babarnew}b, we zoom in at the invariant-mass region
near the $X(4260)$ peak.  We observe that the envelope is flatter than in the
previous set of BABAR data.  This seems to indicate that the reaction
$e^{+}e^{-}\to J/\psi\pi^{+}\pi^{-}$ has a rather constant cross section,
except near thresholds and near resonances of the $c\bar{c}$ propagator.
We indicate the envelope in Fig.~\ref{babarnew} by a solid line that is
almost constant, but having a small bell-shaped contribution which,
for these new data, peaks around 4.35 GeV. Near 4.26 GeV, on top of the smooth
envelope, we follow the data with an eye-guiding line.
Furthermore, just as in Fig.~\ref{babarlow}b, we also show 
the thresholds of the channels $DD^{\ast}$, $D^{\ast}D^{\ast}$,
$D_{s}D_{s}$, $D_{s}D_{s}^{\ast}$, and $D_{s}^{\ast}D_{s}^{\ast}$.
For the latter channel, we display a narrow band representing a possible
spreading of the precise threshold value, in accordance with the different
results for the $D_{s}^{\ast}$ mass reported in
Refs.~\cite{PRL53p2465,PLB156p441,PRL58p2171,PLB207p349,PRD50p1884,PRL75p3232}.
We observe that the new BaBar $X(4260)$ signal seems to be linked to the
opening of the $D_{s}^{\ast}D_{s}^{\ast}$ channel.
However, in view of the above analysis for the further data
of the reaction $e^{+}e^{-}\to J/\psi\pi^{+}\pi^{-}$,
we may not conclude that the peak just corresponds
to the opening of the $D_{s}^{\ast}D_{s}^{\ast}$ channel.
Namely, as we have shown here for various channels,
this would lead to a dip in the production amplitude, not a peak.
However, from the mere fact that the $X(4260)$ signal peaks
at the opening of the $D_{s}^{\ast}D_{s}^{\ast}$ channel,
we may conclude that the $X(4260)$ is neither a resonance
of the $c\bar{c}$ propagator,
nor the consequence of a normal breaking of the $c\bar{c}$ string.
Nevertheless, from its line shape it appears undeniable
that the opening of the $D_{s}^{\ast}D_{s}^{\ast}$ channel
is a relevant factor for its existence \cite{HEPPH0605317}.

The considerations of the preceding paragraph naturally lead to
the following qualitative picture for the $X(4260)$ signal. The
formation of a $D_{s}^{\ast}D_{s}^{\ast}$ pair requires the creation
of $s\bar{s}$.  In the process of strange valence quarks being created,
the $s\bar{s}$ system couples to $f_{0}(980)$, which, in its turn,
couples relatively weakly to pion-pion \cite{NPB320p1,PLB495p300}.
Hence, an intermediate $J/\psi f_{0}(980)$ pair is possible.
The lower-lying thresholds for charm-strange meson pairs
do not have sufficient invariant mass to allow for the formation of
a $J/\psi f_{0}(980)$ pair, with a minimum required energy of
$(4.077\pm0.01)$ GeV. Consequently, $D_{s}^{\ast}D_{s}^{\ast}$
is the first channel where such a process can occur,
taking place at and just above threshold, because at higher energies
the $D_{s}^{\ast}D_{s}^{\ast}$ pair gains too much kinetic energy for
the $s\bar{s}$ pair to stay close to each other during enough time.
The observed signal (see Fig.~\ref{babarnew})
has indeed the characteristics of a threshold opening,
not of some kind of genuine resonance.

In conclusion, we have found further indications supporting
the observation of the $\psi (5S, 4790)$, $\psi (4D, 4870)$,
$\psi (6S, 5130)$, and $\psi (5D, 5290)$
$c\bar{c}$ resonances. Furthermore, we also found here a direct indication
for the existence of the $\psi (3D, 4550)$ $c\bar{c}$ resonance,
which in Ref.~\cite{EPL85p61002} had been deduced
from the treshold behaviour in the reaction
$e^{+}e^{-}\to\Lambda_{c}^{+}\Lambda_{c}^{-}$. Admittedly, our
approach in the present paper is nonstandard, by analysing what we model to
be missing pieces in the data, instead of the data themselves. However, the
remarkable agreement with our recent and totally different analysis of the
BELLE data supports the trustworthiness
of our conclusions. Consequently, we recommend that in experimental analyses
more attention be paid to the structure of any ``background'',
in particular to missing-signal dips.

The shape of the new BABAR $X(4260)$ signal adds to our conviction
\cite{HEPPH0605317}
that the opening of the $D_{s}^{\ast}D_{s}^{\ast}$ channel
plays a crucial role in its dynamics.
Moreover, our present analysis definitely excludes the possibility
that the $X(4260)$ is just a normal $c\bar{c}$ vector state.

Finally, the broad structure which we have suggested 
for the envelope of the data in Fig.~\ref{babar}a,
and also in Fig.~\ref{babarnew}, might be of a true dynamical origin.
Actually, it suggests to consider, next to the $X(4260)$ peak,
a very broad structure with central position at about 4.3--4.35 GeV and
a very large width of several hundreds of MeVs.
A recent detailed three-body calculation for $J/\psi\pi^{+}\pi^{-}$
supports the existence of a very wide structure near 4.3 GeV \cite{EKA}.

\vskip 10pt
We are grateful for the rather precise measurements
of the BABAR collaboration, which made the present analysis possible,
This work was supported in part by
the \emph{Funda\c{c}\~{a}o para a Ci\^{e}ncia e a Tecnologia}
\/of the \emph{Minist\'{e}rio da Ci\^{e}ncia, Tecnologia e Ensino Superior}
\/of Portugal, under contract CERN/\-FP/\-83502/\-2008.

\newcommand{\pubprt}[4]{#1 {\bf #2}, #3 (#4)}
\newcommand{\ertbid}[4]{[Erratum-ibid.~#1 {\bf #2}, #3 (#4)]}
\def\EPL{Europhys.\ Lett.}
\def\NPB{Nucl.\ Phys.\ B}
\def\PLB{Phys.\ Lett.\ B}
\def\PRD{Phys.\ Rev.\ D}
\def\PRL{Phys.\ Rev.\ Lett.}


\begin{thebibliography}{26}
\bibitem{PRL95p142001}
B.~Aubert {\it et al.}  [BABAR Collaboration],
{\it Observation of a broad structure in the $\pi^{+}\pi^{-}J/\psi$
mass spectrum around 4.26-GeV/c$^{2}$},
\pubprt{\PRL}{95}{142001}{2005}
[arXiv:hep-ex/0506081].

\bibitem{ARXIV08081543}
B.~Aubert  [BABAR Collaboration],
{\it Study of the $\pi^{+}\pi^{-}J/\psi$ mass spectrum
via Initial-State Radiation at BaBar},
arXiv:0808.1543 [hep-ex].

\bibitem{PLB667p1}
C.~Amsler {\it et al.} \/[Particle Data Group Collaboration],
{\it Review of Particle Physics},
\pubprt{\PLB}{667}{1}{2008}.

\bibitem{PRL96p162003}
T.~E.~Coan {\it et al.}  [CLEO Collaboration],
{\it Charmonium decays of Y(4260), psi(4160), and psi(4040)},
\pubprt{\PRL}{96}{162003}{2006}
[arXiv:hep-ex/0602034].

\bibitem{PRL99p182004}
C.~Z.~Yuan {\it et al.}  [BELLE Collaboration],
{\it Measurement of $e^{+}e^{-}\to\pi^{+}\pi^{-}J/\psi$ cross section
via initial-state radiation at BELLE},
\pubprt{\PRL}{99}{182004}{2007}
[arXiv:0707.2541 [hep-ex]].

\bibitem{PLB625p212}
S.~L.~Zhu,
{\it The possible interpretations of Y(4260)},
\pubprt{\PLB}{625}{212}{2005}
[arXiv:hep-ph/0507025].

\bibitem{PRD72p031503}
F.~J.~Llanes-Estrada,
{\it Y(4260) and possible charmonium assignment},
\pubprt{\PRD}{72}{031503}{2005}
[arXiv:hep-ph/0507035].

\bibitem{PRD72p054023}
X.~Liu, X.~Q.~Zeng and X.~Q.~Li,
{\it Possible molecular structure of the newly observed Y(4260)},
\pubprt{\PRD}{72}{054023}{2005}
[arXiv:hep-ph/0507177].

\bibitem{PLB631p164}
E.~Kou and O.~Pene,
{\it Suppressed decay into open charm for the Y(4260) being an hybrid},
\pubprt{\PLB}{631}{164}{2005}
[arXiv:hep-ph/0507119].

\bibitem{PRD72p031502}
L.~Maiani, V.~Riquer, F.~Piccinini and A.~D.~Polosa,
{\it Four quark interpretation of Y(4260)},
\pubprt{\PRD}{72}{031502}{2005}
[arXiv:hep-ph/0507062].

\bibitem{HEPPH0605317}
E.~van Beveren and G.~Rupp,
{\it Is the Y(4260) just a coupled-channel signal?},
arXiv:hep-ph/0605317.

\bibitem{ARXIV08111755}
E.~van Beveren and G.~Rupp,
{\it The spectrum of charmonium in the Resonance-Spectrum Expansion},
in Proceedings {\it Bled Workshops in Physics},
Vol.~9, no.~1, pp 26-29 (2008)
[arXiv:0811.1755 [hep-ph]].

\bibitem{OZI}
S.~Okubo,
{\it $\Phi$ meson and unitary symmetry model},
\pubprt{Phys.\ Lett.}{5}{165}{1963};\\
G.~Zweig,
{\it An $SU_{3}$ model for strong interaction symmetry and its breaking},
CERN Reports TH-401 and TH-412 (1963);
see also
{\it Developments in the Quark Theory of Hadrons}, Vol. 1, 22-101 (1981)
editted by D.~B.~Lichtenberg and S.~P.~Rosen;\\
J.~Iizuka, K.~Okada and O.~Shito,
{\it Systematics and phenomenology of boson mass levels (3)},
\pubprt{Prog.\ Theor.\ Phys.}{35}{1061}{1966}.

\bibitem{EPL85p61002}
E.~van~Beveren, X.~Liu, R.~Coimbra, and G.~Rupp,
{\it Possible $\psi(5S)$, $\psi(4D)$, $\psi(6S)$, and
$\psi(5D)$ signals in $\Lambda_{c}\bar{\Lambda}_{c}$},
\pubprt{\EPL}{85}{61002}{2009}
[arXiv:0809.1151 [hep-ph]].

\bibitem{PRD21p772}
E.~van Beveren, C.~Dullemond, and G.~Rupp,
{\it Spectra and strong decays of $c\bar{c}$ and $b\bar{b}$ states},
\pubprt{\PRD}{21}{772}{1980}
\ertbid{\ D}{22}{787}{1980}.

\bibitem{PRD32p189}
S.~Godfrey and N.~Isgur,
{\it Mesons in a relativized quark model with chromodynamics},
\pubprt{\PRD}{32}{189}{1985}.

\bibitem{PRL101p172001}
G.~Pakhlova {\it et al.}  [BELLE Collaboration],
{\it Observation of a near-threshold enhancement in the
$e^{+}e^{-}\to\Lambda_{c}^{+}\Lambda_{c}^{-}$
cross section using initial-state radiation},
\pubprt{\PRL}{101}{172001}{2008}
[arXiv:0807.4458 [hep-ex]].

\bibitem{PRL53p2465}
H.~Aihara {\it et al.}  [TPC Collaboration],
{\it },
\pubprt{\PRL}{53}{2465}{1984}.

\bibitem{PLB156p441}
A.~E.~Asratyan {\it et al.}  [ITEP, SERP Collaboration],
{\it },
\pubprt{\PLB}{156}{441}{1985}.

\bibitem{PRL58p2171}
G.~T.~Blaylock {\it et al.}  [Mark III Collaboration],
{\it },
\pubprt{\PRL}{58}{2171}{1987}.

\bibitem{PLB207p349}
H.~Albrecht {\it et al.}  [ARGUS Collaboration],
{\it },
\pubprt{\PLB}{207}{349}{1988}.

\bibitem{PRD50p1884}
D.~Brown {\it et al.}  [CLEO Collaboration],
{\it },
\pubprt{\PRD}{50}{1884}{1994}.

\bibitem{PRL75p3232}
J.~Gronberg {\it et al.}  [CLEO Collaboration],
{\it },
\pubprt{\PRL}{75}{3232}{1995}.

\bibitem{NPB320p1}
J.~E.~Augustin {\it et al.}  [DM2 Collaboration],
{\it Study of the $J/\Psi$ decay into five pions},
\pubprt{\NPB}{320}{1}{1989}.

\bibitem{PLB495p300}
E.~van Beveren, G.~Rupp and M.~D.~Scadron,
{\it Why is the $f_{0}(9890)$ is mostly $s\bar{s}$},
\pubprt{\PLB}{495}{300}{2000}
\ertbid{\ B}{509}{365}{2001}
[arXiv:hep-ph/0009265].

\bibitem{EKA}
K.~Khemchandani, private communication.
\end{thebibliography}
\end{document}